\documentclass{aa}
\usepackage{psfig}
\usepackage{natbib}
\usepackage{longtable}
\usepackage{graphicx}
\bibpunct{(}{)}{,}{a}{}{,} 
\def\la{\;
\raise0.3ex\hbox{$<$\kern-0.75em\raise-1.1ex\hbox{$\sim$}}\; }
\def\ga{\;
\raise0.3ex\hbox{$>$\kern-0.75em\raise-1.1ex\hbox{$\sim$}}\; }
\newcommand{\etop}{$m_{\rm e}/m_{\rm p}$}
\newcommand{\dmm}{$\Delta\mu/\mu$}
\newcommand{\daa}{$\Delta\alpha/\alpha$}
\newcommand{\kms}{km~s$^{-1}$}

\newcommand{\cmm}{cm$^{-3}$}
\newcommand{\etal}{{et al.}}
\newcommand{\nhhh}{NH$_3$}
\newcommand{\hcIII}{HC$_3$N}
\newcommand{\hcV}{HC$_5$N}
\newcommand{\hcVII}{HC$_7$N}
\newcommand{\DV}{$\Delta RV$}
\newcommand{\Dv}{$\Delta v$}

\newcommand{\VLSR}{$V_{\scriptscriptstyle \rm LSR}$}

\begin{document}
\title{
Limits on the spatial variations of the electron-to-proton mass ratio
in the Galactic plane\thanks{Based
on observations obtained with the Effelsberg 100-m telescope operated by
the Max-Planck Institut f\"ur Radioastronomie on behalf of the Max-Planck-Gesellschaft (Germany),
and with the Medicina 32-m telescope operated by INAF (Italy).  }
}
\author{
S. A. Levshakov\inst{1,2,3} 
\and
D. Reimers\inst{1}
\and
C. Henkel\inst{4,5}
\and
B. Winkel\inst{4}
\and
A. Mignano\inst{6}
\and
M. Centuri\'on\inst{7}
\and
P. Molaro\inst{7,8}
}
\institute{
Hamburger Sternwarte, Universit\"at Hamburg,
Gojenbergsweg 112, D-21029 Hamburg, Germany
\and
Ioffe Physical-Technical Institute,
Polytekhnicheskaya Str. 26, 194021 St.~Petersburg, Russia
\and
St.~Petersburg Electrotechnical University `LETI', Prof. Popov Str. 5,
197376 St.~Petersburg, Russia \\
\email{lev@astro.ioffe.rssi.ru}
\and
Max-Planck-Institut f\"ur Radioastronomie, Auf dem H\"ugel 69, D-53121 Bonn, Germany
\and
Astronomy Department, King Abdulaziz University, P.O.
Box 80203, Jeddah 21589, Saudi Arabia
\and
INAF~-- Istituto di Radio Astronomia, Via P. Gobetti 101, Bologna, Italy
\and
INAF~-- Osservatorio Astronomico di Trieste, Via Tiepolo 11,
I-34131 Trieste, Italy
\and
Centro de Astrofísica, Universidade do Porto, Rua das Estrelas, 4150-762, Porto, Portugal
}
\date{Received 00  ; Accepted 00}
\abstract
{}
{To validate the Einstein equivalence principle (local position invariance)
by limiting the fractional changes in the electron-to-proton mass ratio, $\mu = m_{\rm e}/m_{\rm p}$, 
measured in Galactic plane objects. }
{High resolution spectral observations of dark clouds in the inversion line of \nhhh(1,1) and
pure rotational lines of other molecules (the so-called ammonia method)
were performed at the Medicina 32-m and the Effelsberg 100-m radio telescopes 
to measure the radial velocity offsets,
$\Delta RV = V_{\rm rot} - V_{\rm inv}$,
between the rotational and inversion transitions which have different
sensitivities to the value of $\mu$. }
{In our previous observations (2008-2010), a mean offset 
of $\langle \Delta RV \rangle = 0.027\pm0.010$ \kms\ [$3\sigma$ confidence level (C.L.)] 
was measured. To test for possible hidden errors, we carried out additional
observations of a sample of molecular cores in 2010-2013. 
As a result, a systematic error in the radial velocities
of an amplitude $\sim 0.02$ \kms\ was revealed.   
The averaged offset between the radial velocities of the rotational transitions of
\hcIII(2-1), \hcV(9-8), \hcVII(16-15), \hcVII(21-20), and \hcVII(23-22),
and the  inversion transition of \nhhh(1,1)
$\langle \Delta RV \rangle = 0.003\pm0.018$ \kms\ ($3\sigma$ C.L.).
This value, when interpreted in terms of 
\dmm~$= (\mu_{\rm obs} - \mu_{\rm lab})/\mu_{\rm lab}$,
constraints the $\mu$-variation at the level of
$\Delta \mu/\mu  < 2\times10^{-8}$ ($3\sigma$ C.L.), 
which is the most stringent limit on the fractional changes in $\mu$
based on astronomical observations. 
}
{}

\authorrunning{Levshakov \etal  }

\titlerunning{
Limits on the spatial variations of the electron-to-proton mass ratio
in the Galactic plane}

\maketitle{}

\section{Introduction}
\label{sect-1}

This study is aimed to test whether the dimensionless physical 
constant, the electron-to-proton mass ratio, $\mu =$ \etop, 
is really constant, or whether it varies with space and time.
The latter would imply, at some level, a violation of 
the Einstein equivalence principle (EEP), i.e.,
local position invariance (LPI) and local Lorentz invariance (LLI),
as suggested in a number of unification theories 
(for reviews, see, e.g., Uzan 2011; Liberati 2013).
In particular, 
a changing fine-structure constant, $\alpha = e^2/\hbar c$,  
accompanied by variation in other coupling constants
can be associated with a violation of LLI (Kosteleck\'y \etal\ 2003).
LPI, on the other hand, postulates that the outcome of any local nongravitational
experiment is independent of where and when it is performed, i.e.,
that the fundamental physical laws are space-time invariant.

Experimental validation of EEP is, therefore, one of the pillars of 
the Standard Model (SM) of particle physics
allowing us to probe the applicability limits of the SM
in testing the most accurate theories such as quantum electrodynamics
and/or in searching for new types of interactions.
At the same time, precision limits delivered from such experiments
serve as restrictions for the numerous new theories beyond the SM and can help to distinguish
between them.

Two coupling constants in question 
are of particular interest for astrophysical experiments since
their fractional changes
\dmm~$= (\mu_{\rm obs} - \mu_{\rm lab})/\mu_{\rm lab}$, and
\daa~$= (\alpha_{\rm obs} - \alpha_{\rm lab})/\alpha_{\rm lab}$
can be measured accurately from spectra of Galactic and extragalactic sources.
Because the fine-structure constant sets the scale of the electromagnetic interaction,
experiments in which electrons interact with electromagnetic fields
(e.g., forming in this way atomic and molecular spectra)
can be used to determine $\alpha$.
The electron-to-proton mass ratio is known to define 
the vibrational ($E_{\rm vib} \propto \sqrt{\mu}$) and rotational ($E_{\rm rot} \propto \mu$) 
frequencies of molecular spectra.
On the other hand,
the electron mass $m_{\rm e}$ is related to the vacuum
expectation value of the Higgs field  (i.e., the scale of the weak nuclear force), 
and $m_{\rm p}$~--- to the quantum chromodynamics energy scale 
(i.e., the strong nuclear force).  
Therefore, molecular transitions can be used to determine $\mu$ and thus 
the ratio between weak and strong nuclear forces.

\begin{figure}[t]
\vspace{0.0cm}
\hspace{0.0cm}\psfig{figure=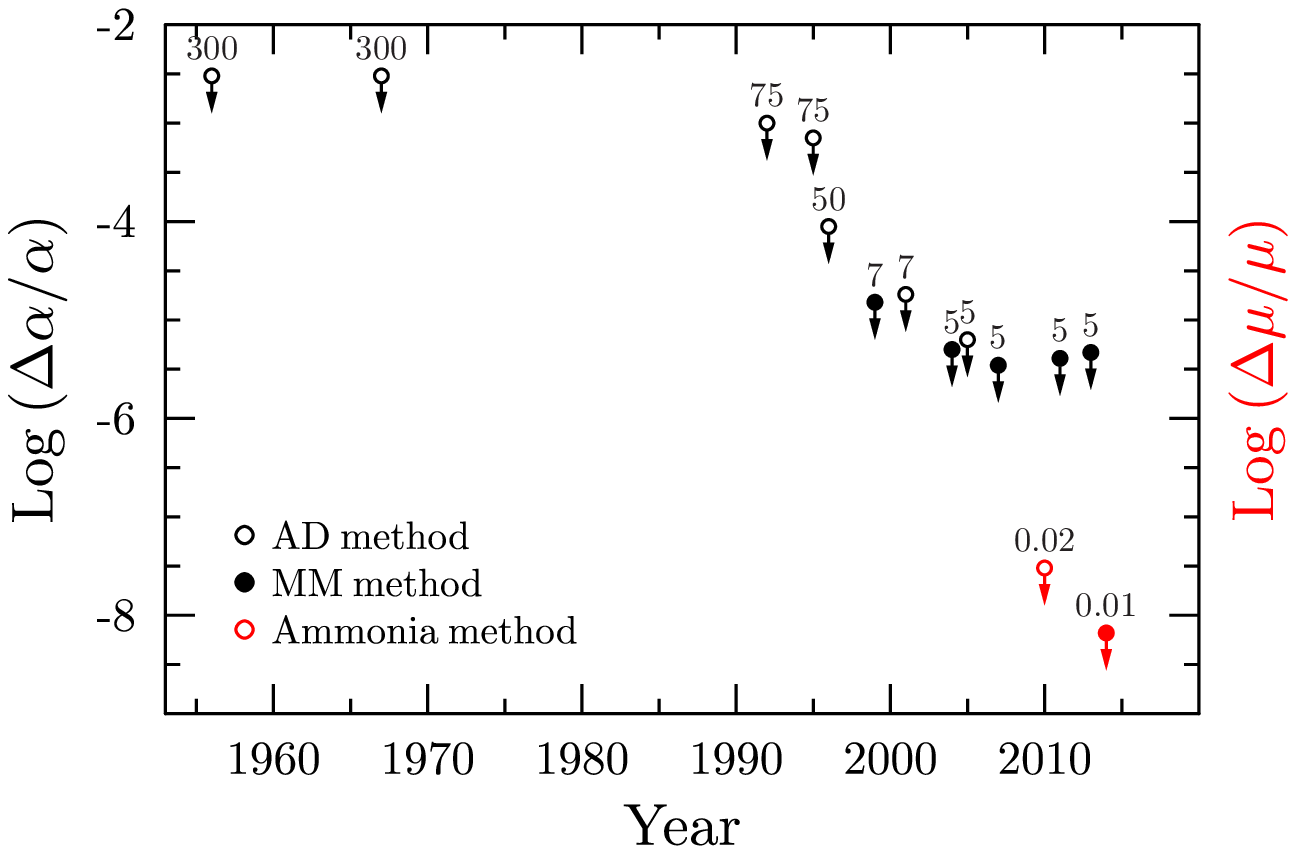,height=8.5cm,width=10.5cm}
\vspace{-2.0cm}
\caption[]{\footnotesize
Astronomical constraints on $\alpha$- and $\mu$-variations ($1\sigma$ C.L.) 
for the period from 1956 to 2013. Above each point, 
the spectral resolution (FWHM in \kms) is indicated.
$\alpha$-variation constraints are based on the alkali-doublet (AD)
and many-multiplet (MM) methods applied to extragalactic targets, whereas $\mu$-variation~-- on the
ammonia method (galactic targets). The data points for the AD method are taken from
Savedoff (1956), Bahcall et al. (1967),
Levshakov (1992), Varshalovich \& Potekhin (1995), Varshalovich et al. (1996),
Murphy et al. (2001), Chand et al. (2005);
MM method~-- Webb et al. (1999), 
Quast et al. (2004), Srianand et al. (2007), Agafonova et al. (2011), 
Molaro et al. (2013); 
ammonia method~-- from Paper~II (red circle), and the current value (red dot). 
The figure shows that limits on the $\alpha$- and $\mu$-variations just follow 
the spectral resolution approximately as
$\Delta\alpha/\alpha\ ({\rm or}\ \Delta\mu/\mu) \propto 1/10{th}$ of the pixel size.
}
\label{fig1}
\end{figure}

In spite of some claims of marginal detections of changes in either $\alpha$ or $\mu$ 
at high redshifts, none have yet been confirmed through independent observations.
As shown in Fig~\ref{fig1}, the most precise constraints on
\daa\ and \dmm\ based on astronomical spectra 
just follow the available spectral resolution.
Up to now, no signals have yet been detected in the range of fractional changes
from $\sim 3\times10^{-2}$ to $\sim 3\times10^{-8}$. Thus, any progress in
improving the existing limits can be achieved from observations 
of narrow spectral lines
involving higher spectral resolutions to resolve completely their profiles.
At the moment, the resolution of radio telescopes exceeds that of optical
facilities by order(s) of magnitude; an additional and very attractive property
of microwave radio observations is that some molecular
transitions from this frequency range are extremely sensitive
to the putative variations of the fundamental physical constants
(for a review see, e.g., Kozlov \& Levshakov 2013).

In 2007, Flambaum \& Kozlov proposed the so-called ammonia method to test the
variability of $\mu$. 
It compares the relative shifts between the inversion transition of \nhhh(1,1)
and pure rotational transitions of other molecules closely tracing the ammonia
spatial distribution. The sensitivity coefficient, $Q_{\rm inv}$, of the inversion
transition to the $\mu$-variation 
is 4.46 times higher than that of the rotational transition ($Q_{\rm rot} = 1$).  
Using this method for a sample of cold molecular cores from the Galactic plane we obtained
the following estimates of the spatial $\mu$-variations: 
\dmm~$= (1.3\pm0.8_{\rm stat}\pm0.3_{\rm sys})\times10^{-8}$
[$1\sigma$ confidence level (C.L.)] 
with the Medicina 32-m telescope (Levshakov et al. 2010a, herein Paper~I), and 
\dmm~$= (2.6\pm0.1_{\rm stat}\pm0.3_{\rm sys})\times10^{-8}$,
based on observations with the Effelsberg 100-m telescope (Levshakov et al. 2010b, herein Paper~II).
The given systematic errors are dominated by uncertainties of
the rest frame frequencies of the observed molecular transitions. 
However, these measurements have been carried out at the highest 
sensitivity available (maximum performance)
and, thus, the true systematic error can be in fact larger because of 
potentially inherent instrumental errors. 
A quantitative assessment of this uncertainty is not a
straightforward procedure and requires in general both a special design of
measurements and a cross-checking of results obtained with different instruments. 
To provide such an assessment we repeated observations of the same
targets employing spectrometers with spectral resolutions,
which are different from those previously used. 
Additionally, we performed a set of measurements at the Effelsberg telescope 
to get a more detailed characteristic of instrumental instabilities which
may affect the values of line radial velocities.
Here we present the obtained results.

\section{Observations}
\label{sect-2}

The following molecular transitions were observed towards 
cold ($T_{\rm kin} \sim 10$K) and dense ($n_{{\rm H}_2} \sim 10^4$ \cmm) starless molecular cores
from the Galactic plane listed in Table~\ref{tbl-1}:
\nhhh\ $(J,K) = (1,1)$ 23.7 GHz, \hcIII\ ($J$ = 2-1) 18.2 GHz, \hcV\ ($J$ = 9-8) 23.9 GHz, 
\hcVII\ ($J$ = 16-15) 18.0 GHz, \hcVII\ ($J$ = 21-20) 23.7 GHz, and \hcVII\ ($J$ = 23-22) 25.9 GHz.
We used the Medicina 32-m and the Effelsberg 100-m radio telescope as described
in Paper~I and II, and in Levshakov \etal\ (2013). 

In November 2010, 
the cores \object{L1512} and \object{L1498} were re-observed
with the 32-m Medicina telescope using the high resolution digital spectrometer MSpec0
with a channel separation of 0.494~kHz (8192 channels). 
This corresponds to $\Delta_{\rm ch} = 0.006$ \kms\ at the position of the ammonia inversion transition \nhhh(1,1). 
For comparison, observations of
the core \object{L1512} in 2008 with the digital spectrometer ARCOS (ARcetri COrrelation Spectrometer)
had $\Delta_{\rm ch} = 0.062$ \kms\ (Paper~I). 
The angular resolutions at the frequencies of the \nhhh(1,1) and 
\hcIII(2-1) lines  were 1.6\arcmin\ and 2.1\arcmin, respectively. 
The observations were performed in the position switching mode (PSW).
The rms pointing uncertainty was $\la 25$\arcsec.

Observations in 2011-2013 with the Effelsberg 100-m telescope targeted a sample of nine
cores from Table~\ref{tbl-1}.
The source coordinates  
were taken from Paper~I and Levshakov \etal\ (2013)  except for \object{L1498A} for 
which they were estimated from Fig.~3 in Kuiper \etal\ (1996).
The source L1498A is a gas condensation within the L1498 dark cloud 
separated by 2\arcmin\ from the \nhhh\ reference position.

In 2011, the measurements were obtained in the frequency switching
(FSW) mode using a frequency throw of $\pm2.5$ MHz. The
backend was a fast Fourier transform spectrometer (FFTS),
operated with a bandwidth of 20 MHz, which simultaneously
provided 16\,384 channels for each polarization. 
The resulting channel width was 
$\Delta_{\rm ch} = 0.015$ \kms. However, the true velocity resolution is
about 1.6 times coarser.

In 2012-2013, we performed the measurements in the PSW mode
with the backend XFFTS (eXtended bandwidth FFTS)
operating with 100\,MHz bandwidth and providing 32\,768 channels
for each polarization. The resulting channel width was $\Delta_{\rm ch} = 0.039$ \kms,
but the true velocity resolution is 1.16 times lower (Klein et al. 2012).

\section{Results}
\label{sect-3}

\begin{figure}[t]
\vspace{0.0cm}
\hspace{0.0cm}\psfig{figure=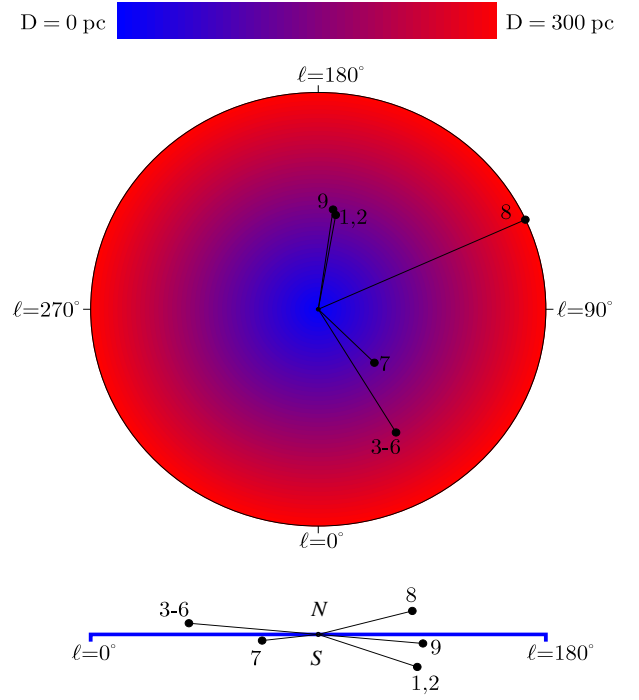,height=12.0cm,width=8.5cm}
\vspace{-1.0cm}
\caption[]{\footnotesize
Schematic location of the observed molecular cores
in projection onto the Galactic plane and equator. 
The Galactic center longitude is $\ell = 0$\degr, by definition.
The distance scale is given by the wedge at the top of the figure.
Numbers mark sources from Table~\ref{tbl-1}.
}
\label{fig2}
\end{figure}

The spatial distribution of the molecular cores in the
galactic coordinates are shown in Fig.~\ref{fig2}.
The cores are located towards both the galactic center and anti-center,
at distances $100$~pc $\la D  \la 300$ pc from the Sun, and
close to the galactic equator, $-19$\degr $< b  < 15$\degr.
The selected sources are known to have narrow molecular emission lines 
(full width at half maximum, FWHM $< 1$ \kms)
what makes them the most suitable targets for precise measurements of
relative radial velocities (RV). At first, we consider results obtained
at Effelsberg, and then at Medicina.

\subsection{Effelsberg observations}
\label{sect-3-1}

To check the reproducibility of the relative RVs of the \nhhh(1,1) and \hcIII(2-1) lines
we re-observed the two molecular cores L1512 and L1498 in 2011.
The procedure was the same as during the 2010 observations:
cores were mapped at the same offsets and in the same lines.
Namely in the (1,1) inversion transition of 
\nhhh\ complemented by rotational lines of
other molecular species.

The line parameters such as the total optical depth in the transition, $\tau$,
the radial velocity, \VLSR, the linewidth, \Dv, 
and the amplitude, $\cal{A}$, were estimated by fitting a one-component
Gaussian model to the observed spectra. 
The model was defined by Eqs.(8-10) in Paper~I.
Since we are mostly interested in the model parameters \VLSR\
and \Dv, their values are listed in Table~\ref{tbl-2}.
Given in parentheses are
the $1\sigma$ errors of \VLSR\ and, directly below, \Dv\ obtained from the diagonal elements of the
covariance matrix calculated for the minimum of $\chi^2$. 
The last column presents the differences \DV~$= V_{\rm rot} - V_{\rm inv}$
between the radial velocities of the rotational lines and the inversion (1,1) line of \nhhh, 
as well as 
their $1\sigma$ errors (numbers in parentheses).
The offsets \DV\ provide a sensitive limit to the variation of $\mu$
(Flambaum \& Kozlov 2007):
$$
{\Delta \mu}/{\mu} = (V_{\rm rot} - V_{\rm inv})/[(Q_{\rm inv} - Q_{\rm rot})c]
\approx 0.3\Delta V/c\ ,
$$
where $c$ is the speed of light, and $Q_{\rm inv}, Q_{\rm rot}$
are the corresponding sensitivity coefficients to changes in $\mu$. 

\begin{figure}[t]
\vspace{0.0cm}
\hspace{-0.7cm}\psfig{figure=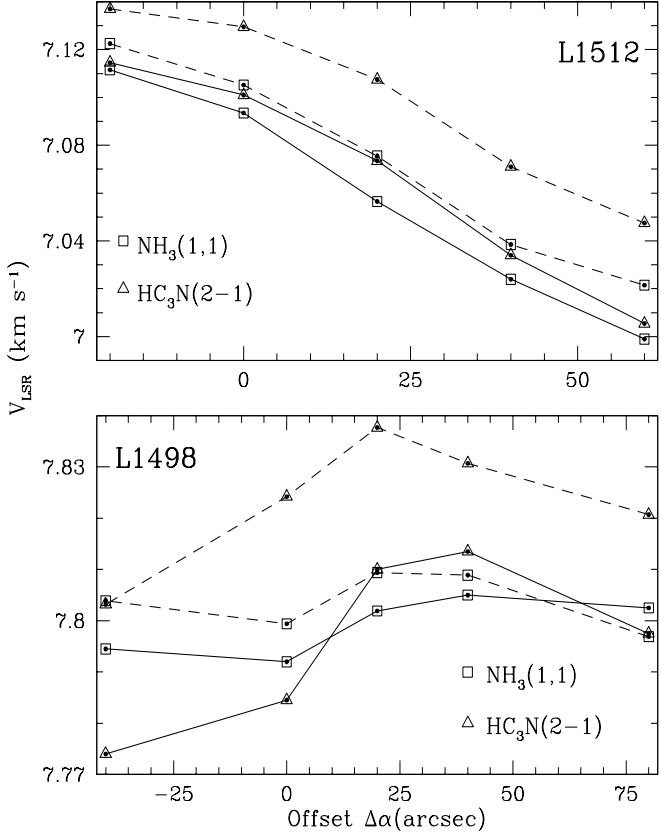,height=12.0cm,width=11.0cm}
\vspace{-0.7cm}
\caption[]{\footnotesize
Line-of-sight velocities ($V_{\rm LSR}$) of
\nhhh\ $(J,K) = (1,1)$ (squares) and \hcIII\ $J = 2-1$ (triangles)
along the main diagonal cuts 
towards the molecular cores L1512 and L1498 
measured in 2010 (dashed lines) and in 2011 (solid lines)
at the Effelsberg 100-m radio telescope.
The half-power beam width at 23 GHz is $40''$ ($50''$ at 18 GHz), the backend
is the fast Fourier transform spectrometer (FFTS) 
with a channel separation $\Delta_{\rm ch} = 0.015$ \kms\ (0.020 \kms\ at 18 GHz). 
}
\label{fig3}
\end{figure}

The RVs measured in 2010 (Paper~II) and 2011
at different radial distances along the main diagonal cuts 
towards L1512 and L1498 are shown in Fig.~\ref{fig3}.
The velocity offsets $\Delta RV$ exhibit
quite a different behavior in these two years:
at the same positions the changes of $\Delta RV$ exceed
considerably its uncertainty interval ($\la 0.003$ \kms, see Figs.~3 and 4
in Paper~II) which is probably caused by unknown systematic errors.

\begin{figure}[t]
\vspace{0.0cm}
\hspace{0.0cm}\psfig{figure=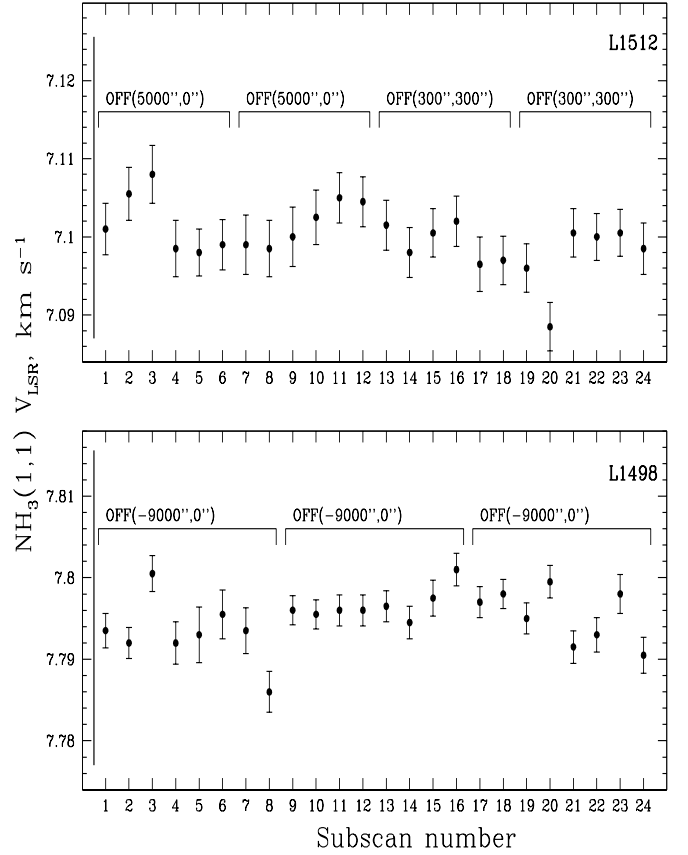,height=12.0cm,width=9.0cm}
\vspace{-0.2cm}
\caption[]{\footnotesize
The line-of-sight velocities ($V_{\rm LSR}$) of the
\nhhh(1,1) transition (dots with $1\sigma$ error bars)
towards the ammonia peaks in the molecular cores L1512 and L1498 
measured continuously in the PSW mode at 
the Effelsberg 100-m radio telescope in April, 2012. 
The exposure time at each point is 150 sec. 
The position switching offsets are shown in parentheses.
The backend was an extended fast Fourier transform spectrometer (XFFTS) 
with a channel separation $\Delta_{\rm ch} = 0.039$ \kms\ (marked by
vertical lines).
Random shifts of the $V_{\rm LSR}$ values are
revealed with an amplitude $\simeq \pm 0.01$ \kms\ 
(i.e., $\simeq 1/4th$ of the channel width).
}
\label{fig4}
\end{figure}

To determine the source of these errors, we performed in 2012 a set of continuous
observations of L1512 and L1498 targeting their ammonia peaks. 
Observing in PSW mode, we also used different OFF positions (for L1512) 
to check possible contamination
from a putative extended background ammonia emission (which was not detected).
The resulting time series are shown in Fig.~\ref{fig4}. The exposure time of each subscan
was 150 sec.
The RV values fluctuate with an amplitude of $\approx \pm 0.01$ \kms, i.e.,
$\approx 1/4th$ of the channel width.

In our observations, the sky frequencies were reset at the onset of each subscan. 
Therefore, the longer a subscan the higher
the error caused by Doppler shifts during the exposure time
(e.g., for a 5 min scan, it is about 0.004 \kms\ at Effelsberg latitude).
We corrected some of our observations to account for residual
Doppler shifts. This did not lead to a significant change in the results, however.

To check whether the sky frequency is identical with the frequency coming out of the
backend we carried out a test with an artificial signal at 22000.78125 MHz.
The synthesizer frequency was accurate to about 1 Hz, and
the frequency scale was found to be accurate to about 32 Hz ($\approx 0.0004$ \kms).

\begin{figure}[t]
\vspace{0.0cm}
\hspace{0.0cm}\psfig{figure=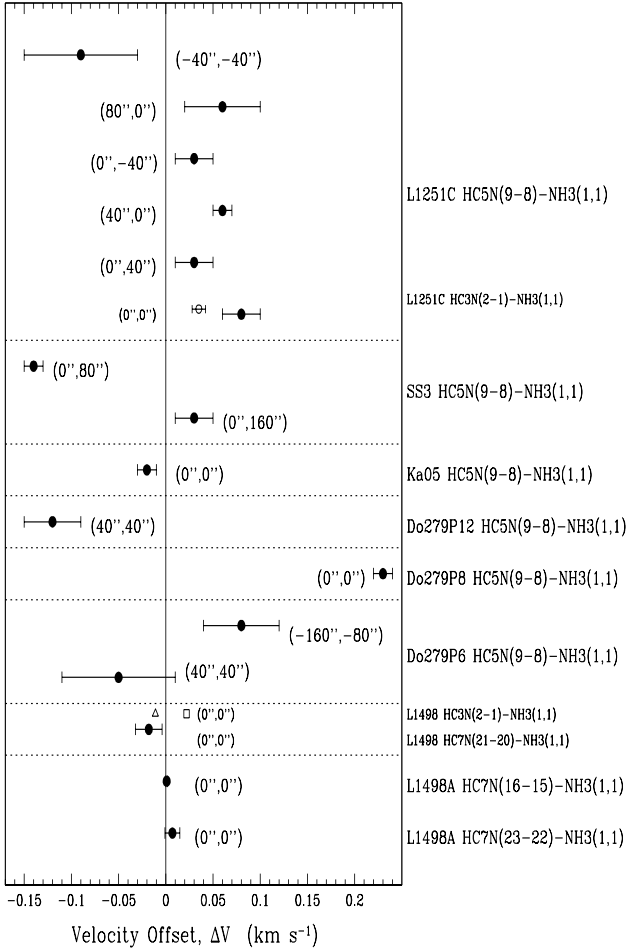,height=13.0cm,width=9.0cm}
\vspace{-0.3cm}
\caption[]{\footnotesize
Radial velocity differences, $\Delta RV$, between rotational 
transitions of different molecules and the 
\nhhh(1,1) inversion line 
for the sources observed at the Effelsberg 100-m telescope (2011-2013). 
1$\sigma$ statistical errors are indicated.
In the panel, given in parentheses are the coordinate offsets in arcsec.
Filled circles~-- this paper;
open triangle, square, and circle~-- Paper~I, II. 
}
\label{fig5}
\end{figure}

Another source of errors which can affect the $\Delta\mu/\mu$
estimates with the ammonia method is
the possible segregation of molecules in molecular cores
giving rise to systematic shifts of RVs.
Figure~\ref{fig5} shows the \DV\ offsets measured 
in 2013 (filled circles)
towards eight molecular cores at different positions indicated in column 2 of Table~\ref{tbl-2}
and columns 3 and 4 of Table~\ref{tbl-1}.
Previous \DV\ values, obtained in 2009-2011, are marked by open symbols.
A spread of the velocity offsets $\Delta RV$ is clearly seen.

Thus, we conclude that noise in the $\Delta RV$ values consists of at least two 
components.  One is due to chemical differentiation and velocity gradients within
the molecular cores, 
possibly being amplified by small variations in the telescope pointing.
The other may originate from
the different optical depths of the hyperfine structure transitions
leading to some scatter in the RVs. 
However, all these effects may act in opposite directions
from one observation to another, and,
being averaged over a sample of targets, should be reduced to some extent.
Applied to our sample shown in Fig.~\ref{fig5} ($n=19$ independent offsets \DV), 
this gives the weighted mean
$\langle \Delta RV \rangle = 0.003 \pm 0.018$ \kms\ ($3\sigma$ C.L.).
Being interpreted in terms of
\dmm\ $= (\mu_{\rm obs} - \mu_{\rm lab})/\mu_{\rm lab}$, 
this value of $\langle \Delta RV \rangle$
constraints the $\mu$-variation at the level of
$\Delta \mu/\mu  < 2\times10^{-8}$ ($3\sigma$ C.L.),
which is the most stringent limit on the spatial variation of $\mu$
based on radio astronomical observations. 

We note in passing that
mapping of the dense molecular cores in different molecular lines shows 
that there is, in general, a good
correlation between \nhhh, N$_2$H$^+$, and \hcIII\ distributions
(Fuller \& Myers 1993; Hotzel et al. 2004; Tafalla et al. 2004; Pagani et al. 2009). 
However, in some clouds \nhhh\ is not traced by \hcIII, as, e.g., in the dark cloud
TMC-1, where peaks of emission lines are offset by $7'$ (Olano et al. 1988).
In our case, we observe systematic velocity shifts between \nhhh\ and other species.
This is expected since C-bearing molecules are usually
distributed in the outer layers of the cores rich in depletion-sensitive molecules,
whereas N-bearing molecules, which are depletion-resistant species, 
trace the inner cores
(examples are given in, e.g., Tafalla et al. 2006).

\subsection{Medicina observations}
\label{sect-3-2}

\begin{figure}[t]
\vspace{0.0cm}
\hspace{0.0cm}\psfig{figure=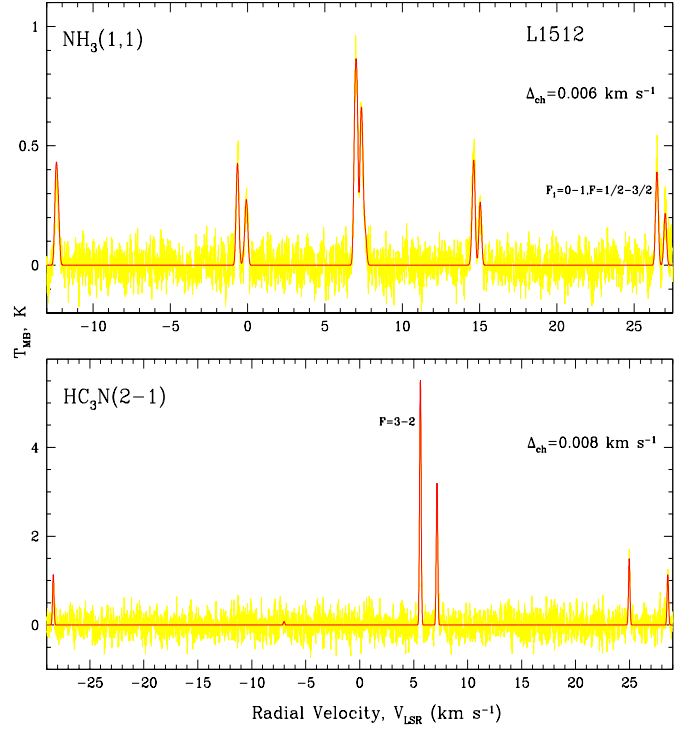,height=10.0cm,width=9cm}
\vspace{0.0cm}
\caption[]{\footnotesize
Spectra of \nhhh(1,1) and \hcIII(2-1) (yellow)
towards the core \object{L1512} (offset $\Delta\alpha, \Delta\delta =$ 0\arcsec,0\arcsec) 
obtained at the Medicina 32-m radio telescope with the high resolution spectrometer MSpec0
(channel widths $\Delta_{\rm ch} = 0.006$ \kms\ and 0.008 \kms\ at 23 GHz and 18 GHz, respectively).
The red curves are the best fit model spectra.
The observed linewidths are \Dv(\nhhh)~$= 0.19$ \kms, and 
\Dv(\hcIII)~$=0.16$ \kms.
}
\label{fig6}
\end{figure}

\begin{figure}[t]
\vspace{0.0cm}
\hspace{0.0cm}\psfig{figure=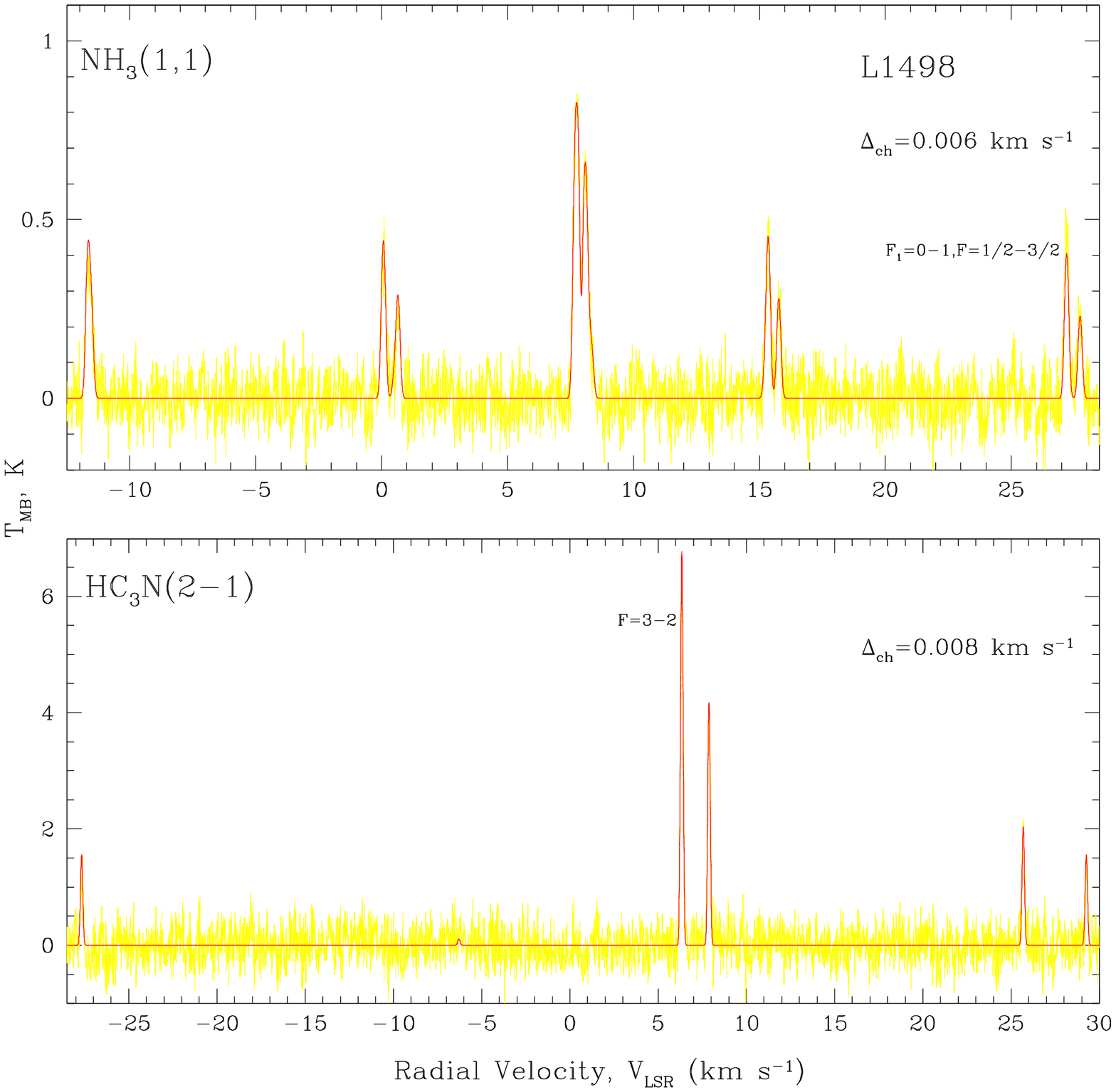,height=10.0cm,width=9cm}
\vspace{0.0cm}
\caption[]{\footnotesize
Same as Fig.~\ref{fig6} but for the core
\object{L1498}. 
The observed linewidths are \Dv(\nhhh)~$= 0.19$ \kms, and 
\Dv(\hcIII)~$=0.16$ \kms.
}
\label{fig7}
\end{figure}

The \nhhh(1,1) and \hcIII(2-1) lines towards the cores \object{L1512} and \object{L1498} are
among the narrowest molecular lines known in the interstellar medium.
The observed FWHM linewidths of the completely resolved hyperfine components, 
the $F'_1 \rightarrow F_1, F' \rightarrow F = 0 \rightarrow 1, 1/2 \rightarrow 3/2$ line
of \nhhh(1,1) and
the $F' \rightarrow F = 3 \rightarrow 2$ transition of \hcIII(2-1) 
are
\Dv~$=0.19$ \kms, and $0.16$ \kms, respectively. The \nhhh\ and \hcIII\ spectra are
shown in Figs.~\ref{fig6} and \ref{fig7}.
The hyperfine components show no kinematic sub-structure and consist of an
apparently symmetric peak profile without broadened line wings or self-absorption
features. In the cores \object{L1512} and \object{L1498}, 
the \nhhh\ hyperfine satellite lines with $\Delta F_1 \neq 0$ are
optically thin, $\tau < 1$, whereas the main transitions with
$\Delta F_1 = 0$  are slightly saturated, $\tau \ga 1$ (Paper~II). 
However, we did not find any significant differences between the RVs of the main
and satellite lines (see Tables~2 and 3 in Paper~II). 
The optical depth of the strongest hyperfine transition $F' \rightarrow F = 3 \rightarrow 2$ 
of \hcIII(2-1) is 0.67 (\object{L1498}) and 0.35 (\object{L1512}). 

If the two molecular transitions are thermally broadened and trace the same material,
then the lighter molecule \nhhh\ should have a wider linewidth as compared to
the heavier molecule \hcIII. This is precisely what is seen in Figs.~\ref{fig6} and \ref{fig7}. 
Thus the observed spectra from \object{L1512} and \object{L1498} 
are consistent with thermally dominated line broadening at a kinetic temperature $T_{\rm kin} \sim 13$~K.
This may indicate that the two species are sampling the same gas.

Table~\ref{tbl-2} lists
the measured offsets of the \nhhh(1,1) and \hcIII(2-1) lines 
as well as their linewidths (observations with MSpec0). 
The measured RVs give a velocity offset \DV\ = $0.001\pm0.009$ \kms\ 
for \object{L1512}, and 
\DV\ = $-0.001\pm0.012$ \kms\ for \object{L1498}, with the mean
$\langle \Delta RV \rangle = 0.000 \pm 0.009$ \kms.
The previous observations of \object{L1512} with ARCOS (Paper~I) 
gave the offset \DV\ = $0.018\pm0.012$ \kms\ 
which is slightly biased with respect to the current estimate (indicated are $3\sigma$ C.L.).

Formally, the new observations show an error of the mean $\langle \Delta RV \rangle $ 
two times lower than that of the Effelsberg result. 
However, since only one single position was observed in both cores, we cannot evaluate
the systematics of MSpec0 (like we do at Effelsberg) and on the basis of the
Effelsberg experience we adopt the same limit,
\dmm~$< 2\times10^{-8}$ ($3\sigma$ C.L.), 
i.e., we conservatively assign a bias of $\sim 0.02$ \kms\ to the possible systematic error of the
Medicina dataset.

\section{Limits on \daa }
\label{sect-4}

The obtained local constraint on the spatial $\mu$-variation, 
$\Delta\mu/\mu < 2\times10^{-8}$ ($3\sigma$ C.L.),
can be used to set limits on changes in $\alpha$.  This, however, is strongly model dependent.  
For example, within the grand unification theory (GUT) a variation of $\alpha$ would imply considerably 
larger fractional changes in the mass scale of the strong force in QCD, $\Lambda_{\rm QCD}$, 
and in quark and electron masses leading to 
$$
\Delta \mu/\mu \sim R \Delta\alpha/\alpha\ ,
$$
where $R \sim 40$ (e.g., Langacker et al. 2002; Flambaum et al. 2004).
This gives a limit on 
$| \Delta\alpha/\alpha | < 10^{-10}$ ($1\sigma$ C.L.). 
A direct estimate of changes in $\alpha$ in the Milky Way, based on observations
of CH sources, sets a $1\sigma$ upper limit $|\Delta\alpha/\alpha| < 10^{-7}$ (Truppe et al. 2013). 

At higher redshifts, the most stringent constraint on cosmological $\mu$-variation
was set at $z = 0.89$, $| \Delta \mu/\mu | < 10^{-7}$  (Bagdonaite et al. 2013).
For $R \sim 40$, this would imply that $| \Delta\alpha/\alpha | < 2.5\times10^{-9}$ ($1\sigma$ C.L.)
at epoch $7\times10^9$ yr, meaning in turn that 
$| \dot{\alpha}/\alpha | < 4\times10^{-19}$ yr$^{-1}$. 
At very high redshift, $z = 5.2$ (epoch 12.9 Gyr), 
the current limit is \daa\ $< 8\times10^{-6}$ ($1\sigma$ C.L.)
corresponding to $|\dot{\alpha}/\alpha| < 6\times10^{-16}$ yr$^{-1}$
(Levshakov \etal\ 2012).
We recall that most stringent limits set by the Oklo fossil reactor 
and by terrestrial atomic clock experiments are, respectively,
$|\dot{\alpha}/\alpha| < 5\times10^{-17}$ yr$^{-1}$ (Uzan 2011), and
$|\dot{\alpha}/\alpha| < 4\times10^{-17}$ yr$^{-1}$ (Rosenband et al. 2008). 

As mentioned above, the value of $R$ is poorly constrained.
Depending on the theory, it varies from $-235$ to +46 (see, e.g., Section~5.3.1 in Uzan 2011).
The only way to distinguish between theories is to measure \daa\ and \dmm\ independently.

\section{ Conclusions}
\label{sect-5}

We have used the Medicina 32-m telescope and 
the Effelsberg 100-m telescope to observe the
\nhhh(1,1) 23.7 GHz, 
\hcIII(2-1) 18.2 GHz, \hcV(9-8) 23.9 GHz, \hcVII(16-15) 18.0 GHz, \hcVII(21-20) 23.7 GHz, and \hcVII(23-22) 25.9 GHz 
spectral lines in molecular cores devoid of associated IR sources.
The principle of local position invariance (LPI) was tested 
in the solar vicinity ($D \la 300$ pc) by comparing the electron-to-proton mass ratio $\mu$
in different physical environments of high terrestrial ($n \sim 10^{19}$ \cmm)
and low interstellar ($n \sim 10^{4}$ \cmm)
densities of baryonic matter. 
No statistically significant changes in $\mu$ have been detected 
at a level of $\sim 10^{-8}$. 

The main results obtained are as follow.
\begin{enumerate}
\item[1.] In order to test the reproducibility of the measurements of the relative
radial velocities between the
\nhhh(1,1) and \hcIII(2-1) transitions observed towards dark molecular cores in 2008-2010
at the Medicina 32-m and
the Effelsberg 100-m telescopes, we re-observed two clouds, L1512 and L1498, and revealed
discrepancies between the $\Delta RV = V_{\rm LSR}({\rm HC}_3{\rm N}) - V_{\rm LSR}({\rm NH}_3)$
values which are as large as the channel width, $\Delta RV \la 0.02$ \kms.
\item[2.] Continuous observations of L1512 and L1498 at Effelsberg
in 2012 at a fixed position towards
the ammonia peaks showed that the measured radial velocity $V_{\rm LSR}({\rm NH}_3)$ fluctuates
during the exposure time of 2 hours with an amplitude $\simeq \pm 0.01$ \kms, i.e.,
with approximately $1/4th$ of the channel width
which does not allow us to measure
radial velocities with uncertainties less than 0.01 \kms.
\item[3.] Tests with the synthesizer frequency at 
2000.78125 MHz showed that the sky frequency is accurate to about 32 Hz, i.e.,
$\approx 0.0004$ \kms\ at the Effelsberg 100-m telescope.
\item[4.] Taking into account the revealed errors and averaging relative velocities over
a sample of eight molecular cores observed in 2013, we find a null offset
$\langle \Delta RV \rangle = 0.003 \pm 0.018$ \kms\ ($3\sigma$ C.L.) 
between the rotational and inversion transitions of the above mentioned molecules
observed with the Effelsberg 100-m telescope.
If this offset is interpreted in terms of
\dmm\ $= (\mu_{\rm obs} - \mu_{\rm lab})/\mu_{\rm lab}$, then 
the spatial $\mu$-variation is constrained at the level of
$\Delta \mu/\mu  < 2\times10^{-8}$ ($3\sigma$ C.L.), 
what is the strictest limit for the validity of the LPI principle
based on radio astronomical observations. 
\item[5.] A similar null offset was found from high spectral resolution
observations of two cores \object{L1512} and 
\object{L1498} with the Medicina 32-m telescope in 2010:
$\langle \Delta RV \rangle = 0.000 \pm 0.009$ \kms\ ($3\sigma$ C.L.).
\end{enumerate}

\begin{acknowledgements}
We thank the staff of the Effelsberg 100-m telescope and the Medicina 32-m telescope
for the assistance in observations
and acknowledge the help of Giuseppe Maccaferri.
SAL's work is supported by the grant DFG
Sonderforschungsbereich SFB 676 Teilprojekt C4, and in part
by Research Program OFN-17 of the Division of Physics, 
Russian Academy of Sciences.
\end{acknowledgements}

\bibliographystyle{aa}

\clearpage

\clearpage
\begin{table*}[t!]
\centering
\caption{Source positions and radial velocities}
\label{tbl-1}
\begin{tabular}{c l  l r  r  l  c}
\hline
\hline
\noalign{\smallskip}
No. & \multicolumn{1}{c}{Core} & \multicolumn{2}{c}{Position} &  \multicolumn{1}{c}{$V_{\scriptscriptstyle\rm LSR}$}
& \multicolumn{1}{c}{Other name} & \multicolumn{1}{c}{Telescope} \\
&  & \multicolumn{1}{c}{$\alpha_{2000}$} & \multicolumn{1}{c}{$\delta_{2000}$} & \multicolumn{1}{c}{(\kms)} 
&  & \multicolumn{1}{c}{used}\\
&  &  ($^{\rm h}$: $^{\rm m}$: $^{\rm s}$) & ({\degr}: {\arcmin}: {\arcsec}) \\
\noalign{\smallskip}
\hline

\noalign{\smallskip}
1 & \object{L1498}    & 04:10:51 & +25:09:58 & 7.81    & & 32-m / 100-m\\
2 & \object{L1498A}   & 04:10:58 & +25:09:15 & 7.75    & & 100-m \\
3 & \object{Do279P6} & 18:29:07 & +00:30:51   & 7.75 & \object{Ser G3-G6}$^1$ & 100-m \\
4 & \object{Do279P12}& 18:29:47 & +01:14:53   & 7.59 & \object{BDN~31.57+5.37}$^2$  & 100-m \\
5 & \object{SS3}      & 18:29:57 & $-$02:00:10 & 7.39 & & 100-m \\
6 & \object{Do279P8} & 18:30:04 & $-$02:48:14 & 8.52 & & 100-m \\
7 & \object{Ka05}   &   19:37:01 & +07:34:30   & 8.40 & \object{B335}$^3$, \object{CB199}$^4$ & 100-m \\
8 & \object{L1251C}   & 22:35:54 & +75:18:55 &  $-4.73$ &  &  100-m \\
9 & \object{L1512}   & 05:04:10 & +32:43:09 &  7.11 &  &  32-m / 100-m\\

\noalign{\smallskip}
\hline
\noalign{\smallskip}
\multicolumn{7}{l}{{\bf References.} (1) Cohen \& Kuhi 1979; (2) Dobashi \etal\ 2005; (3) Barnard 1927;} \\
\multicolumn{7}{l}{(4) Clemens and Barvainis 1988. }
\end{tabular}
\end{table*}

\begin{table*}[t!]
\centering
\caption{Radial velocities, $V_{\scriptscriptstyle\rm LSR}$, linewidths (FWHM), $\Delta v$, and velocity offsets, 
$\Delta RV = V_{\rm rot} - V_{\rm inv}$,
measured with the Effelsberg 100-m and the Medicina 32-m telescopes.
}
\label{tbl-2}
\begin{tabular}{l r@{,}l  l l l l l l r}
\hline
\hline
\noalign{\smallskip}
\multicolumn{1}{c}{Source} & \multicolumn{2}{c}{Offset} & 
\multicolumn{6}{c}{$V_{\scriptscriptstyle\rm LSR}$, \kms /$\Delta v$, \kms } & 
\multicolumn{1}{c}{$\Delta RV$,} \\
 & {$\Delta\alpha$} & {$\Delta\delta$} & 
NH$_3$(1,1) & HC$_7$N(23-22) & HC$_7$N(16-15) & HC$_7$N(21-20) &  HC$_5$N(9-8) & HC$_3$N(2-1) &
\multicolumn{1}{c}{\kms }\\
 & ({\arcsec})&({\arcsec})  \\
\noalign{\smallskip}
\hline

\noalign{\smallskip}

\multicolumn{10}{l}{\it Effelsberg:} \\

\noalign{\smallskip}
\object{L1498A} &   0&0   & 7.753(1)$^a$ & 7.760(2)$^a$  &  & $\ldots$ & $\ldots$ & $\ldots$ & 0.007(2) \\[-2pt]
& \multicolumn{2}{c}{}   & 0.204(3) & 0.096(7) &   \\[-2pt]
& \multicolumn{2}{c}{}   &          &       & 7.753(6)$^a$ & $\ldots$ & $\ldots$ & $\ldots$ & 0.000(6)\\[-2pt]   
& \multicolumn{2}{c}{}   &          &       & 0.13(1)  &  \\[2pt]   
\object{L1498} &  0&0    & 7.798(2)$^a$ & $\ldots$ & $\ldots$ & 7.78(1)$^a$ & $\ldots$ & $\ldots$ & $-0.02(1)$ \\[-2pt]
& \multicolumn{2}{c}{}   & 0.186(5) &  &  & 0.08(2) &  \\[2pt]
\object{Do279P6}& 40&40  & 8.045(7)$^c$ & $\ldots$ & $\ldots$ & $\ldots$ & 8.00(6)$^a$ & $\ldots$ & $-0.05(6)$ \\[-2pt]
& \multicolumn{2}{c}{}   & 0.63(1)  &  &  &  &  0.04(3)  \\[2pt]
&          $-160$&$-80$  & 7.69(1)$^c$  & $\ldots$ & $\ldots$ & $\ldots$ & 7.77(4)$^a$  &  $\ldots$ & 0.08(4) \\[-2pt]
& \multicolumn{2}{c}{}   & 0.96(3)  &  &  &  & 0.5(1)  \\[2pt]
\object{Do279P8} & 0&0   & 8.524(7)$^c$ & $\ldots$ & $\ldots$ & $\ldots$ &  8.75(1)$^a$ & $\ldots$ & 0.23(1) \\[-2pt]
& \multicolumn{2}{c}{}   & 0.26(2)  &  &  &  & 0.13(2) & \\[2pt]
\object{Do279P12} & 40&40& 8.290(8)$^c$ & $\ldots$ & $\ldots$ & $\ldots$ & 8.17(3)$^a$ & $\ldots$ & $-0.12(3)$ \\[-2pt]
& \multicolumn{2}{c}{}   & 0.91(2)  &  &  &  & 0.23(5) & \\[2pt]
\object{Ka05} &  0&0     & 8.401(4)$^b$ & $\ldots$ & $\ldots$ & $\ldots$ & 8.38(1)$^a$ & $\ldots$ & $-0.02(1)$ \\[-2pt]
& \multicolumn{2}{c}{}   & 0.365(8) &  &  &  & 0.32(8) & \\[2pt]
\object{SS3} & 0&160     & 7.622(6)$^b$ & $\ldots$ & $\ldots$ & $\ldots$ & 7.65(2)$^a$ &  $\ldots$ & 0.03(2) \\[-2pt]
& \multicolumn{2}{c}{}   & 0.396(3) &  &  &  & 0.50(3) & \\
             & 0&80      & 7.668(2)$^b$  & $\ldots$ & $\ldots$ & $\ldots$ & 7.53(1)$^a$ & $\ldots$ & $-0.14(1)$ \\[-2pt]
& \multicolumn{2}{c}{}   & 0.490(5) &  &  &  & 0.38(3) & \\[2pt]
\object{L1251C} & 0&0  & \multicolumn{1}{l}{$-4.722(3)^a$} & $\ldots$ & $\ldots$ & $\ldots$ & 
\multicolumn{1}{l}{$-4.64(2)^a$} & $\ldots$ & 0.08(2) \\[-2pt]
& \multicolumn{2}{c}{}   & 0.290(7) &  &  &  & 0.33(3) & \\
               & 0&40  &\multicolumn{1}{l}{$-4.616(9)^a$} & $\ldots$ & $\ldots$ & $\ldots$ & $-4.59(2)^a$ & $\ldots$ & 0.03(2)\\[-2pt]
& \multicolumn{2}{c}{}   & 0.30(2) &  &  &  & 0.28(3) & \\
               & 40&0 &\multicolumn{1}{l}{$-4.733(5)^a$}& $\ldots$ & $\ldots$ & $\ldots$ & $-4.67(1)^a$ & $\ldots$ & 0.06(1)\\[-2pt]
& \multicolumn{2}{c}{}   & 0.28(1) &  &  &  & 0.27(3) & \\
            & $-40$&0& \multicolumn{1}{l}{$-4.718(6)^a$} & $\ldots$ & $\ldots$ & $\ldots$ & $-4.69(2)^a$ & $\ldots$ & 0.03(2)\\[-2pt]
& \multicolumn{2}{c}{}   & 0.29(1) &  &  &  & 0.30(8) & \\
           & 80&0 & \multicolumn{1}{l}{$-4.755(6)^a$}  & $\ldots$ & $\ldots$ & $\ldots$ & $-4.70(4)^a$& $\ldots$ & 0.06(4) \\[-2pt]
& \multicolumn{2}{c}{}   & 0.29(1) &  &  &  & 0.37(7) & \\
        &$-40$&$-40$ & \multicolumn{1}{l}{$-4.805(6)^a$} & $\ldots$ & $\ldots$ & $\ldots$ & $-4.89(6)^a$ & $\ldots$ & $-0.09(6)$ \\[-2pt]
& \multicolumn{2}{c}{}   & 0.35(1) &  &  &  & 0.28(2) & \\
\noalign{\smallskip}

\multicolumn{10}{l}{\it Medicina:} \\
\noalign{\smallskip}

\object{L1512} &  0&0    & 7.160(2)$^d$ & $\ldots$ & $\ldots$ & $\ldots$ & $\ldots$ & 7.161(2)$^d$ & 0.001(3) \\[-2pt]
& \multicolumn{2}{c}{}   & 0.191(7) &  &  & & & 0.16(1) &  \\

\object{L1498} &  0&0    & 7.883(3)$^d$ & $\ldots$ & $\ldots$ & $\ldots$ & $\ldots$ & 7.882(2)$^d$ & $-0.001$(4) \\[-2pt]
& \multicolumn{2}{c}{}   & 0.185(8) &  &  & & & 0.16(1) &  \\

\noalign{\smallskip}
\hline
\noalign{\smallskip}
\multicolumn{10}{l}{{\bf Notes.}  Observations: $^a$2013, $^b$2012, $^c$2011, $^d$2010. Rest frame frequencies: 
\nhhh(1,1) 23694.495487(60) MHz, }\\
\multicolumn{9}{l}{\hcVII(23-22) 25943.8549(7) MHz, 
\hcVII(21-20) 23687.8974(6) MHz, \hcVII(16-15) 18047.9697(5) MHz,}\\
\multicolumn{10}{l}{\hcV(9-8) 23963.9007(1) MHz, \hcIII(2-1) 18196.2169(2) MHz. 
Notations: upper values are \VLSR\ and the lower ones~-- $\Delta v$;}\\
\multicolumn{10}{l}{the values in parenthesis give $1\sigma$ standard deviations, referring to the last given digit of the
respective parameter.}
\end{tabular}
\end{table*}

\end{document}